\documentclass[9pt,twocolumn,twoside]{osajnl}

\journal{ol} 
\usepackage{bm}

\setboolean{shortarticle}{true} 

\newcommand{\KB}{\textcolor{black}}

\title{Spin-Hall effect of light at a tilted polarizer}

\author[1,2,*]{K. Y. Bliokh}
\author[3,4]{C. Prajapati}
\author[3,5]{C. T. Samlan}
\author[3]{N. K. Viswanathan}
\author[1,6]{F. Nori}

\affil[1]{Theoretical Quantum Physics Laboratory, RIKEN Cluster for Pioneering Research, Wako-shi, Saitama 351-0198, Japan}
\affil[2]{Nonlinear Physics Centre, The Australian National University, RSPE, Canberra, ACT 0200, Australia}
\affil[3]{School of Physics, University of Hyderabad, Hyderabad 500046, India}
\affil[4]{Department of Physics, Koneru Lakshmaiah Education Foundation, Hyderabad 500075, India}
\affil[5]{Department of Engineering Science, Institute for Advanced Science, The University of Electro-Communications, Chofu 182-8585, Japan}
\affil[6]{Physics Department, University of Michigan, Ann Arbor, Michigan 48109-1040, USA}

\affil[*]{Corresponding author: kostiantyn.bliokh@riken.jp}

\ociscodes{(260.0260) Physical optics; (260.5430) Polarization; (240.5440) Polarization-selective devices}

\begin{abstract}
We describe the spin-Hall effect of light (as well as the angular Goos-H\"{a}nchen effect) at a tilted linear-dichroic plate, such as a usual linear polarizer. Although the spin-Hall effect at a tilted polarizer was previous associated with the {\it geometric} spin-Hall effect of light (which was contrasted to the regular spin-Hall effect) [J. Korger  {\it et al.}, Phys. Rev. Lett. {\bf 112}, 113902 (2014)], we show that the effect is actually an example of the {\it regular} spin-Hall effect that occurs at tilted anisotropic plates [K. Y. Bliokh {\it et al.}, Optica {\bf 3}, 1039 (2016)]. Moreover, our approach reveals the {\it angular} spin-Hall shift, which is absent in the ``geometric'' approach. We verify our theory experimentally using the method of quantum weak measurements. 
\end{abstract}

\setboolean{displaycopyright}{true}

\begin{document}

\maketitle

\vspace{0.5cm}

Spin-Hall effect of light (SHEL) is one of the main manifestations of optical spin-orbit interactions, which have been intensively studied in the past decade
 \cite{Onoda2004, Bliokh2006, Bliokh2007, Hosten2008, Aiello2008, Fedoseyev2008, Bliokh2009, Merano2010, Fedoseyev2013} (for reviews, see \cite{Bliokh2013, Bliokh2015, Ling2017}).
This phenomenon typically appears at reflection or transmission of light beams at various interfaces, and it produces shifts of the right- and left-hand circularly polarized beams in opposite directions, orthogonally to the plane of propagation. Note that these shifts generically appear in both real (position) and momentum (direction) spaces. 

The regular SHEL at a planar interface originates from the interference of individual plane-wave components in the beam, which propagate in slightly different directions and acquire slightly different complex reflection or transmission coefficients \cite{Bliokh2007,Bliokh2013}. A convenient theoretical description of the SHEL is provided by a quantum-like formalism with the generalized wavevector-dependent Jones-matrix operators of the interface and expectation values of the position and momentum of light \cite{Bliokh2013,Dennis2012,Gotte2012,Toppel2013}. Such description also unifies the transverse SHEL shifts (also known as the Imbert--Fedorov shifts in the case of the Fresnel reflection/refraction) and longitudinal (in-plane) beam shifts associated with the Goos--H\"{a}nchen (GH) effect \cite{Aiello2008,Bliokh2013,Dennis2012,Gotte2012,Toppel2013}.

In 2009, Aiello {\it et al}. put forward the concept of ``{\it geometric} SHEL'' \cite{Aiello2009}. This is a spin-dependent transverse shift of the centroid of the energy flux density (Poynting vector) through a tilted cross-section of a paraxial optical beam {\it in free space}. This remarkable effect also occurs for vortex beams carrying orbital angular momentum \cite{Bekshaev2009,Kong2012}, as well as for relativistically Lorentz-transformed beams (i.e., for space-time tilted cross-sections) \cite{Bliokh2012,Smirnova2018}.
The geometric SHEL was eventually observed experimentally as a SHEL of a beam transmitted through a {\it tilted \KB{dichroic} polarizer} \cite{Korger2014}. \KB{In fact, the authors of Ref.~\cite{Korger2014} claimed that they observed ``a novel kind of the geometric SHEL'', different from that in Ref.~\cite{Aiello2009}. Nonetheless,} it was contrasted to the regular SHEL at optical interfaces because ``the geometric SHEL is practically independent of Snell's law and the Fresnel formulas for the interface'', and because of its $\tan\theta$ dependence on the angle of incidence, which is ``in striking contrast to the typical $\cot\theta$ angular dependence of the conventional SHEL'' \cite{Korger2014}.

Recently, we showed, both theoretically and experimentally, that the SHEL appears not only at isotropic Snell--Fresnel interfaces, but also at the transmission of light beams through {\it anisotropic} wave plates \cite{Bliokh2016} \KB{(see also \cite{GP1,GP2})}. The general formalism for the SHEL \cite{Bliokh2013,Dennis2012,Gotte2012,Toppel2013} is perfectly applicable in this case, with the Fresnel coefficients being substituted by the polarization-dependent transmission coefficients of the plate.

Here we show that the spin-Hall effect for a beam transmitted through a tilted polarizer is an example of the {\it regular} SHEL for a tilted anisotropic plate. The only difference is that in the polarizer case, the birefringence considered in \cite{Bliokh2016} \KB{(i.e., polarization-dependent phase of the transmitted wave)} is changed to dichroism \KB{(i.e., polarization-dependent amplitude of the transmitted wave)}. Moreover, the $\tan\theta$ behaviour observed in \cite{Korger2014} is perfectly consistent with the typical $\cot\theta$ behaviour of the conventional SHEL, because the angle $\theta$ for anisotropic plates should be counted with respect to the {\it anisotropy axis} rather than the normal to the plate \cite{Bliokh2016}. 
Importantly, we describe and observe the {\it angular} SHEL shift, inherent in the standard SHEL theory but {\it absent} in the geometric-SHEL approach.
We verify our theory by experimental measurements of the SHEL via the method of ``quantum weak measurements'' \cite{Hosten2008, Qin2009, Zhou2012, Gorodetski2012, Dennis2012, Gotte2012, Toppel2013, Jayaswal2014, Goswami2014, Bliokh2016}.

Although the geometric SHEL interpretation advocated in \cite{Korger2014} could also be  relevant to the tilted-polarizer system, we emphasize that the geometric SHEL strongly depends on the choice of the theoretical quantity under consideration. Namely, calculating the centroid of either the energy density, or energy flux density, or momentum flux density, etc., results in very different shifts, varying from zero to a double-SHEL shift \cite{Aiello2009,Bliokh2012,Smirnova2018,Wang2018}. Accordingly, the measured effect also depends on the sensitivity of the detector to this or that quantity, and it is not clear if the tilted polarizer is sensitive to the energy flux density as it was assumed in \cite{Korger2014}. In contrast, our ``regular SHEL'' approach uses only the transmission coefficients of the polarizer and directly provides the intensity distribution of the transmitted beam. These quantities are unambiguous in both theory and experiment.


We start with the theoretical description of the problem, using the nomenclature of the closely-related work \cite{Bliokh2016} (see also \cite{Bliokh2013,Dennis2012,Gotte2012,Toppel2013}). The geometry of the problem is shown in Fig.~1(a). The polarization of the incident $z$-propagating paraxial beam is described by the normalized Jones vector $\left| \psi  \right\rangle = {\begin{pmatrix} E_x \\ E_y 
\end{pmatrix}}$, $\left\langle \psi \! \right.\left|\, \psi \right\rangle = |E_x|^2 + |E_y|^2 =1$. The polarizer is tilted in the $(x,z)$ plane, such that its absorption axis forms an angle $-\theta$ with the $z$-axis, and it transmits mostly the $y$-polarization. In this geometry, and in the zero-order approximation of the incident {\it plane-wave} field, the dichroic action of the polarizer can be described by the Jones matrix  
%
\begin{eqnarray}
{{\hat M}_0} = {\begin{pmatrix}
{{T_x ( \theta )}} & 0 \\
0&{{T_y ( \theta )}}
\end{pmatrix}} ,
\label{E1}
\end{eqnarray}
%
so that the Jones vector of the transmitted wave is $\left| \psi' \right\rangle = \hat{M}_0 \left| \psi \right\rangle$ (throughout this paper, the prime indicates the transmitted-beam properties). Here $T_{x,y}$ are the amplitude transmission coefficients for the $x$- and $y$-polarized waves, which can depend on $\theta$. For an ideal polarizer, $T_x = 0$ and $T_y = 1$, but for real dichroic plates we can assume $|T_x/T_y| \ll 1$. Note also that $T_{x,y} = \exp (\mp i\Phi/2)$ corresponds to the problem with a birefringent waveplate described in \cite{Bliokh2016}. 

In the first-order paraxial approximation, taking into account that the beam consists of many plane waves \KB{\cite{FS1984}} with their wavevector directions described by small angles ${\bm \Theta} = \left(\Theta_x,\Theta_y \right)\simeq \left(k_x/k,k_y/k \right)$ [see Fig.~1(a)], the Jones matrix (\ref{E1}) acqures $\bm \Theta$-dependent corrections \cite{Bliokh2013,Dennis2012,Gotte2012,Toppel2013} and can be wrtten as \cite{Bliokh2016}:
%
\begin{eqnarray}
\hat M\left( {\bm \Theta}  \right) = \left( {\begin{matrix}
{{T_x}\left( {1 + {\Theta _x}{{\cal X}_x}} \right)}&{{T_x}\,{\Theta _y}{{\cal Y}_x}}\\
{ - {T_y}\,{\Theta _y}{{\cal Y}_y}}&{{T_y}\left( {1 + {\Theta _x}{{\cal X}_y}} \right)}
\end{matrix}} \right),
\label{E2}
\end{eqnarray}
%
where 
%
\begin{eqnarray}
{{\cal X}_{x,y}} = \frac{{d \ln T_{x,y} }}{{d\theta }}~, \qquad
{{\cal Y}_{x,y}} = \left( {1 - \frac{T_{y,x}}{T_{x,y}}} \right) \! \cot \theta~, 
\label{E3}
\end{eqnarray}
%
are the typical GH and SHEL terms. 

\begin{figure}[t]
\centering
\includegraphics[width=\linewidth]{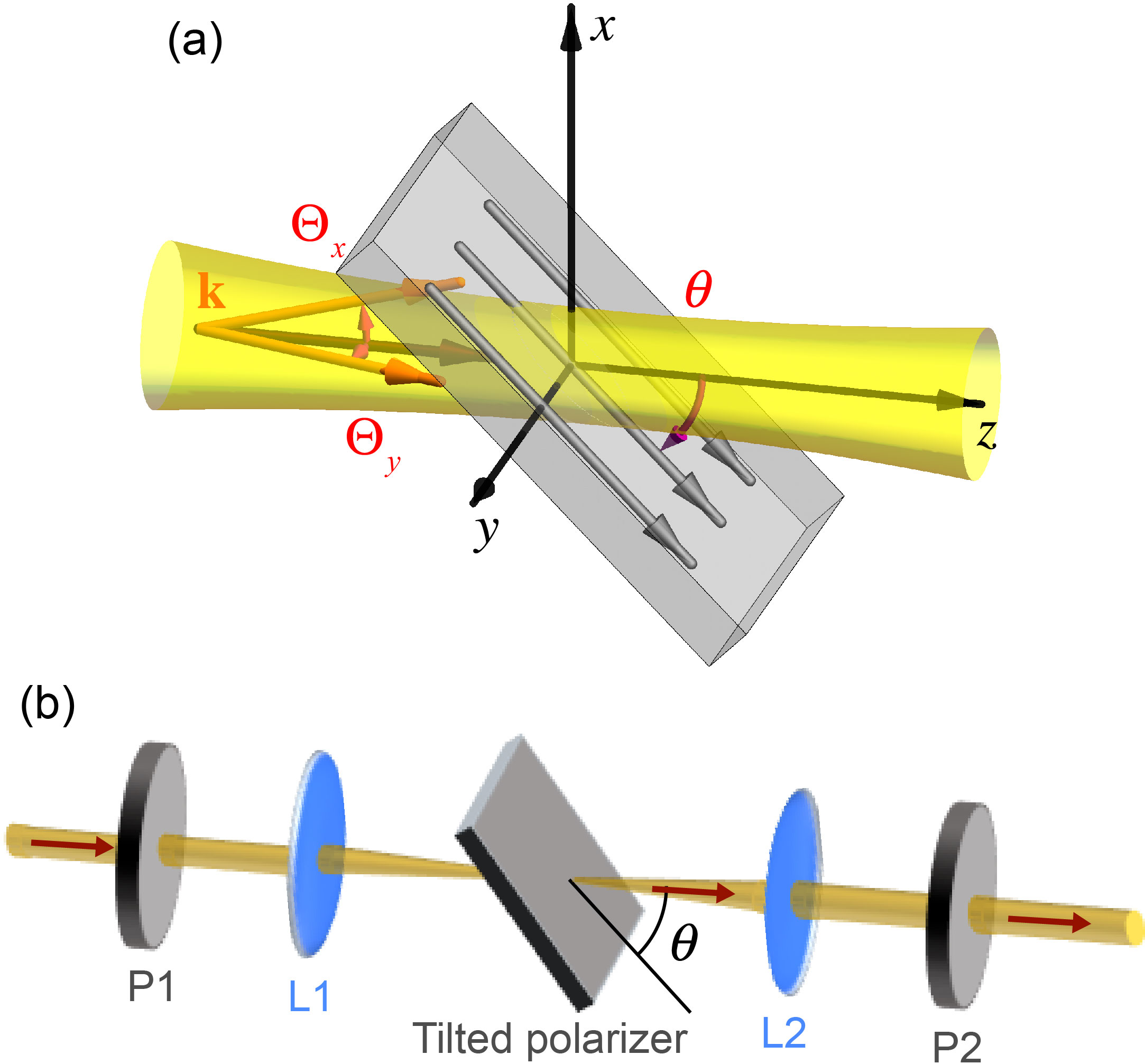}
\caption{(a) Schematic of the transmission of a paraxial beam through a tilted polarizer. The absorption axis of the poalrizer lies in the $(x,z)$ plane at an angle $-\theta$ with respect to the $z$ axis. The small angles ${\bm \Theta} = \left(\Theta_x,\Theta_y \right)$ determine the directions of the wave vectors ${\bf k}$ in the incident beam. (b) Schematic of the experimental setup used for the quantum weak measurements of the spin-Hall shift of the transmitted beam.}
\label{F1}
\end{figure}

Writing the Jones matrix (\ref{E2}) as $\hat{M} = \left(1 - i\,k\, \Theta_x \hat{X} - i\,k\, \Theta_y \hat{Y} \right) \hat{M}_0$, we obtain the operators of the GH and SHEL shifts $\hat{X}$ and $\hat{Y}$ \cite{Dennis2012,Gotte2012,Toppel2013,Bliokh2016}. These operators are generally non-Hermitian; the real and imaginary parts of their expectation values correspond to the spatial (position) and angular (direction) shifts of the transmitted beam, respectively. Assuming, for simplicty, that ${\rm Im} T_{x,y} = 0$, we find that the expectation value of $\hat{X}$ is purely imaginary, and it describes the {\it angular GH shift} \cite{Merano2009,Gotte2013} of the transmitted beam:
%
\begin{eqnarray}
\left\langle {\Theta_x^\prime } \right\rangle  = 
\frac{1}{z_R}{\rm Im} \frac{{\left\langle {\psi '} \right|\hat X\left| {\psi '} \right\rangle }}{{\left\langle {\psi '}\! \right.\left|\, {\psi '} \right\rangle }} =  
\frac{1}{{k\,{z_R}}}\frac{{d\ln {\mathcal T}}}{{d\theta }}~.
\label{E4}
\end{eqnarray}
%
Here, $z_R$ is the Rayleigh distance (the angular shift is counted from the focal plane, so that the actual shift of the beam is $\langle X_z^\prime \rangle = z \left\langle {\Theta _x^\prime } \right\rangle$, where $z$ is the propagation distance from the focal plane), and ${\mathcal T}^2 = {\left\langle \psi ' \! \right.\left|\, \psi ' \right\rangle} = 
{|T_x E_x|^2 + |T_y E_y|^2}$ is the intensity transmission coefficient (${\mathcal T}=|T_{x,y}|$ for the $x$- and $y$-polarized beams). Equation (\ref{E4}) perfectly agrees with the angular GH expression for the beam reflection/refraction at planar interfaces \cite{Aiello2008,Bliokh2009,Bliokh2013}.

The expectation value of the SHEL operator $\hat{Y}$ is generally complex:
%
\begin{eqnarray}
\left\langle \hat{Y}' \right\rangle \equiv
\frac{{\left\langle {\psi '} \right|\hat Y\left| {\psi '} \right\rangle }}{{\left\langle {\psi '}\! \right.\left| \,{\psi '} \right\rangle }} =  \!\left[ - \frac{\sigma }{{2k}}\frac{{{{\left( {{T_x} - {T_y}} \right)}^2}}}{{{{\cal T}^2}}} 
+ i\frac{\chi }{{2k}}\frac{{T_x^2 - T_y^2}}{{{{\cal T}^2}}}\right]\! \cot\theta~.
\label{E5}
\end{eqnarray}
%
Here, $\sigma = 2\,{\rm Im}(E_x^*E_y)$ and $\chi = 2\,{\rm Re}(E_x^*E_y)$ are the third and the second Stokes parameters of the incident beam, which describe the degrees of the circular (spin) and diagonal linear polarizations, respectively. The real part of this expression describes the spatial SHEL shift in the focal plane of the beam: $\langle Y' \rangle = {\rm Re} \langle \hat{Y}' \rangle$, while, similarly to Eq.~(\ref{E4}), its imaginary part describes the angular SHEL shift $\langle \Theta_y^{\prime} \rangle = z_R^{-1}{\rm Im} \langle \hat{Y}' \rangle$. The total beam shift at a distance $z$ from the focal plane is given by $\langle Y_z^\prime \rangle = \langle Y' \rangle + z \langle \Theta_y^{\prime} \rangle$.

Remarkably, Eq.~(\ref{E5}) {\it precisely coincides} with the well-known expression for the spatial and angular SHEL shifts for the beam refraction at a Snell--Fresnel interface, if we set the refraction angle to be equal to the angle of incidence, $\theta ' = \theta$ \cite{Bliokh2006,Bliokh2007,Bliokh2009,Bliokh2013}. Importantly, in contrast to the angular GH shift (\ref{E4}), the SHEL shifts (\ref{E5}) are present in the {\it ideal-polarizer} case, $T_x = 0$ and $T_y = 1$. Then, expression (\ref{E5}) yields 
$\left\langle \hat{Y}' \right\rangle = -(\sigma + i \chi)\cot\theta \big/ (2k|E_y|^2)$. For a circularly-polarized incident beam, $\sigma=\pm 1$, $\chi=0$, $|E_y|^2 = 1/2$, this equation yields the expression $-\sigma \cot\theta /k$ for the {\it geometric SHEL} (taking into account that $\theta$ in \cite{Aiello2009,Korger2014} corresponds to $\theta - \pi/2$ in this paper). Thus, the regular SHEL formalism, with equations completely equivalent to those at planar Snell--Fresnel interfaces, perfectly explain the SHEL in the tilted-polarizer system. However, our theory also predicts the {\it angular} SHEL shift given by ${\rm Im} \left\langle \hat{Y}' \right\rangle$, which is {\it absent} in the geometric-SHEL approach.

To prove the validity of our approach and to measure the SHEL at a tilted polarizer, we employ the method of {\it quantum weak measurements} \cite{Hosten2008, Qin2009, Zhou2012, Gorodetski2012, Dennis2012, Gotte2012, Toppel2013, Jayaswal2014, Goswami2014, Bliokh2016}. In this approach, the incident beam is linearly $y$-polarized (``pre-selected''), $\left| \psi  \right\rangle = {\begin{pmatrix} 0 \\ 1 \end{pmatrix}}$, then it passes (almost freely) through the tilted polarizer under consideration, and then another polarizer projects (``post-selects'') the transmitted beam onto an almost-orthogonal polarization state $\left| \varphi  \right\rangle \simeq {\begin{pmatrix} 1 \\ \varepsilon \end{pmatrix}}$, $|\varepsilon|\ll 1$. The SHEL shifts of the resulting beam after the ``post-selection'' are described by the complex {\it weak value}: 
%
\begin{eqnarray}
{\left\langle Y' \right\rangle _w} = \frac{{\left\langle \varphi  \right|\hat Y\left| {\psi '} \right\rangle }}{{\left\langle \varphi  \right.\! \left| \,{\psi '} \right\rangle }} =  - \frac{i}{{{\varepsilon ^*}k}}\left( {1 - \frac{{{T_x}}}{{{T_y}}}} \right)\cot \theta 
= - \frac{i}{{{\varepsilon ^*}k}}  {\mathcal Y}_y~.
\label{E6}
\end{eqnarray}
%
The small $\varepsilon$ in the denominator makes the shift (\ref{E6}) much larger than the typical subwavelength shift (\ref{E5}). This is the desired enhancement from the weak-measurement method. Furthermore, choosing $\varepsilon$ to be real or imaginary, one can choose between the angular and spatial nature of the shift (\ref{E6}) \cite{Gorodetski2012}. The angular shift is easier to measure due to additional amplification from the propagation factor $z/z_R \gg 1$ \cite{Hosten2008,Aiello2008,Qin2009,Bliokh2016}, it is absent in the geometric-SHEL approach, and therefore we choose $\varepsilon$ to be real.

\begin{figure}[t]
\centering
\includegraphics[width=\linewidth]{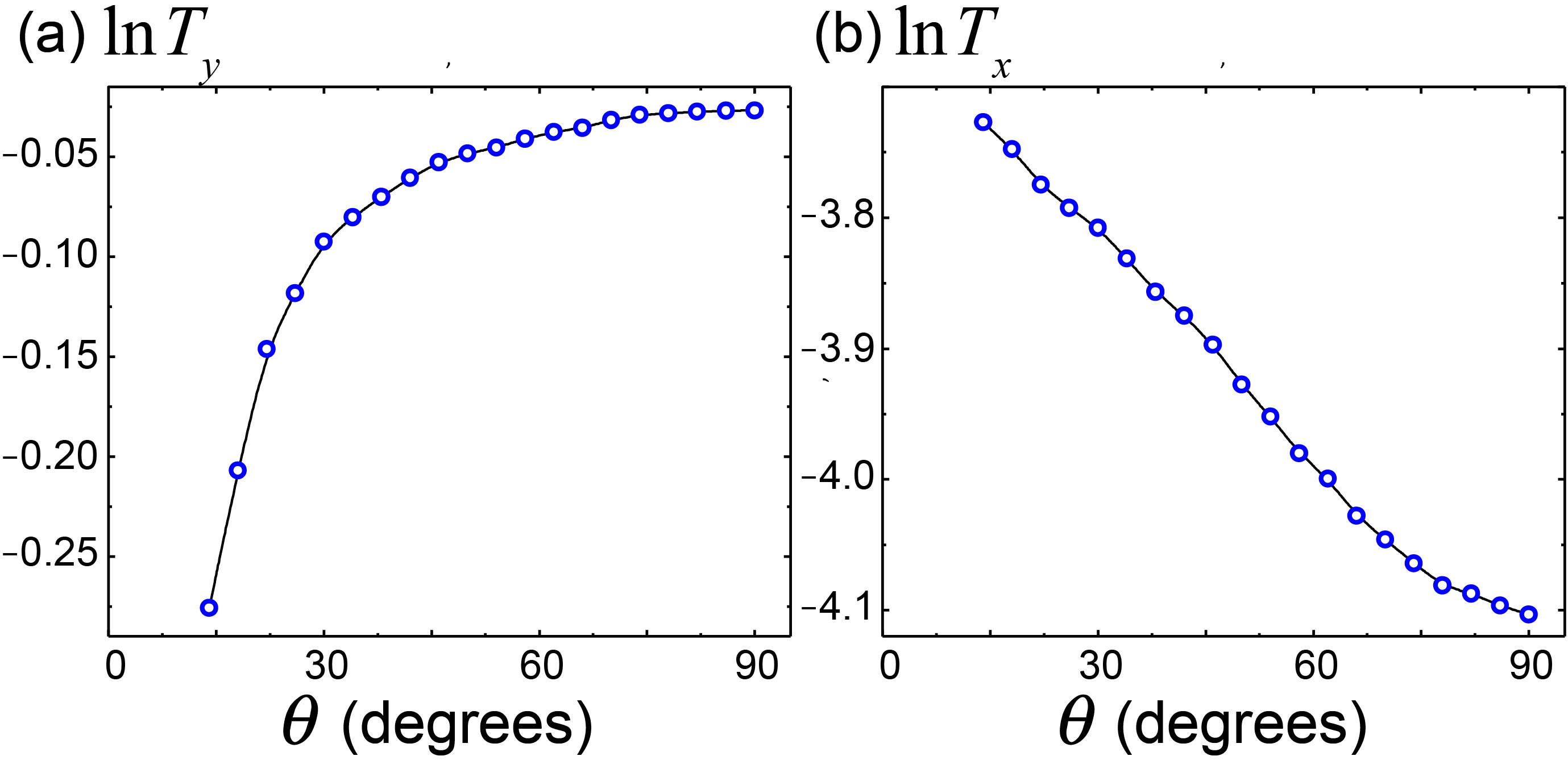}
\caption{Experimentally measured angular dependencies of the logarithms of the transmission coefficients $T_{x,y}(\theta)$, Eq.~(\ref{E1}). The derivatives of these dependencies determine the angular GH shifts according to Eqs.~(\ref{E3}) and (\ref{E4}).}
\label{F2}
\end{figure}

 Note that the weak value (\ref{E6}) diverges at $\varepsilon \to 0$ or $\theta \to 0$. This divergence can be regularized by taking into account the {\it nonlinear} correction in the weak-measurement approach \cite{Gorodetski2012,Kofman2012,Dennis2013,Asano2016}. 
This regularization is universal for Gaussian beams and realized by introducing the prefactor $\left(1 + \dfrac{k}{2z_R} \left| {\left\langle Y' \right\rangle _w} \right|^2 \right)^{-1}$ in all the beam shifts \cite{Asano2016}. As a result, for purely real $\varepsilon$, the directly observed beam shift is:
%
\begin{eqnarray}
\left\langle {Y_z^\prime} \right\rangle  
= \frac{z}{z_R}\,\frac{{{\rm Im} {{\left\langle {Y'} \right\rangle }_w}}}{{1 + \dfrac{k}{{2{z_R}}}{{\left| {{{\left\langle {Y'} \right\rangle }_w}} \right|}^2}}}
\simeq  - \frac{z}{z_R}\,\frac{{\cot \theta }}{{\varepsilon\, k}}\left({{1 + \dfrac{{{{\cot }^2}\theta }}{{2{\varepsilon ^2}k\,z_R}}}}\right)^{-1}\!,
\label{E7}
\end{eqnarray}
%
where we set $T_x/T_y \simeq 0$, because in our experiment this quantity is $\sim 10^{-3}$, see Fig.~2.

\begin{figure}[t]
\centering
\includegraphics[width=\linewidth]{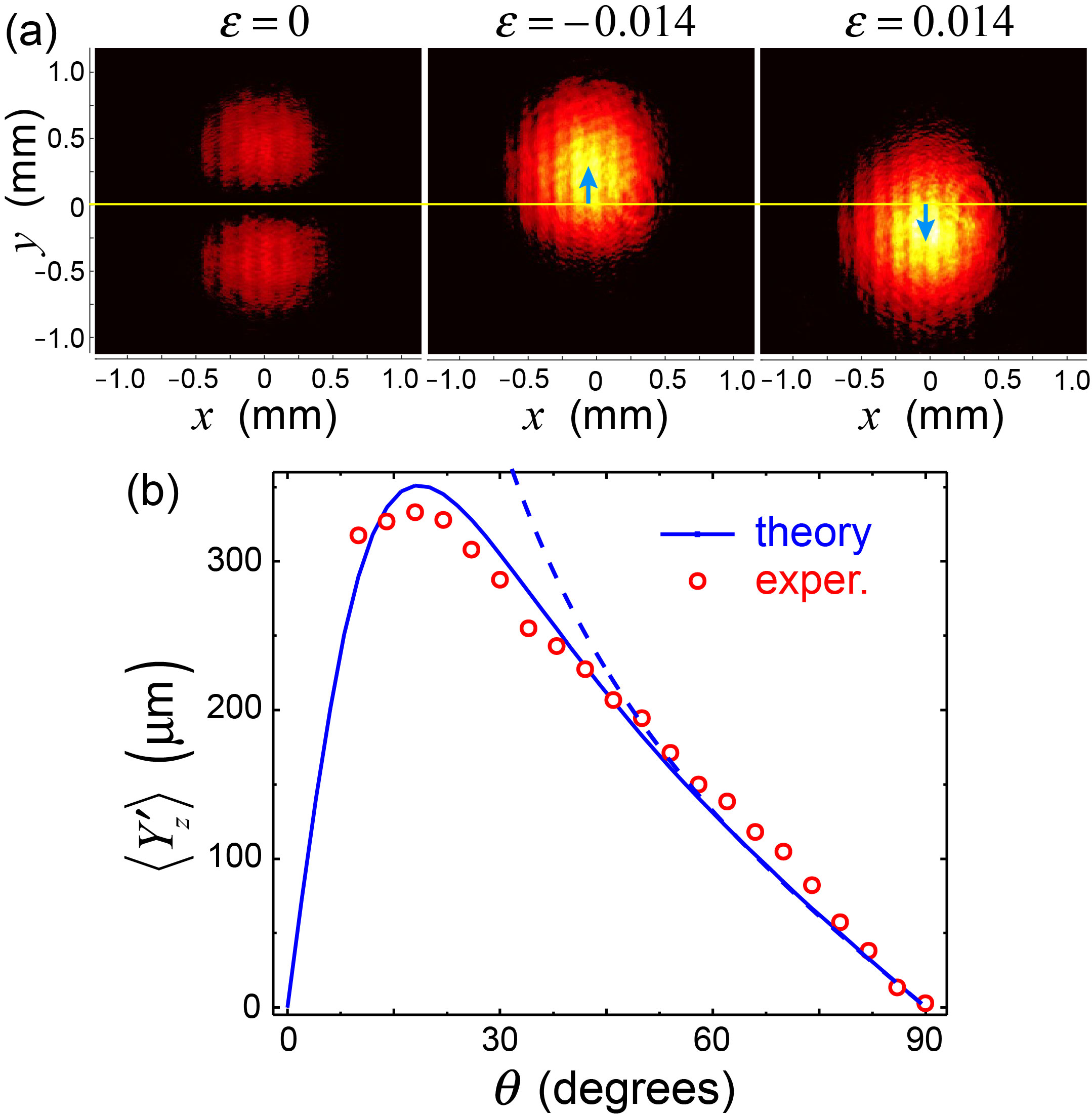}
\caption{Spin-Hall effect of light at a tilted polarizer. (a) Experimentally measured transverse intensity distributions of the transmitted beam at different angles $\varepsilon$ of the polarizer P2, which determines the ``post-selection'' in the weak-measurement scheme Fig.~1(b). (b) Experimentally measured angular dependence of the spin-Hall shift vs. the theoretical dependence (\ref{E7}). \KB{The experimental shift was calculated as a half-difference of the shifts for $\varepsilon = - 0.014$ and $\varepsilon = + 0.014$, to make it independent of the choice of the $y=0$ line.} The dashed curve corresponds to the theoretical dependence (\ref{E7}) without the nonlinear weak-measurement correction.}
\label{F3}
\end{figure}

For experimental measurements of the predicted effect, we used a sheet polarizer (Thorlabs, USA) and an intensity-frequency stabilized He-Ne laser (Thorlabs, USA) with the wavelength $\lambda = 2\pi/k = 0.6328~\mu{\rm m}$. To characterize the polarizer parameters quantitatively, we measured the dependencies of the transmission coefficients $T_{x,y}$ of the $x$- and $y$-polarized colimated laser beams through the $y$-oriented polarizer on the angle of the polarizer tilt, $\theta$. These dependencies (in the logarithmic scale) are shown in Fig.~2. Their derivatives determine the angular GH effect, Eqs.~(\ref{E3}) and (\ref{E4}), but for the SHEL measurements it is sufficient to use the approximation of an ideal polarizer: $T_x/T_y \simeq 0$.

To measure the SHEL at a tilted polarizer, we used the experimental setup schematically shown in Fig.~1(b) and entirely analogous to that used in Ref. \cite{Bliokh2016} with a tilted waveplate. Polarizers P1 and P2 produced pre-selected and post-selected polarization states $\left| \psi  \right\rangle$ and $\left| \varphi \right\rangle$, respectively, while the lenses controlled the propagation factor $z/z_R$. Namely, the first lens, L1, of focal length 50~mm, produced a focused Gaussian beam with the Rayleigh range $z_R = k w_0^2 /2 \simeq 2.273~{\rm mm}$ (determined by measuring the beam waist at the focal plane, $w_0 \simeq 21.4~\mu{\rm m}$), while the second lens, L2, of focal length 75~mm, collimated the beam and provided the effective propagation distance $z = 75~{\rm mm}$. 

First, by tuning the small post-selection parameter $\varepsilon$ (via rotation of the polarizer P2), at the fixed angle $\theta = 42$~deg, we observed typical SHEL deformations of the intensity distribution in the transmitted beam, Fig.~3(a). Namely, the two-hump Hermite-Gaussian $y$-distribution takes place for $\varepsilon=0$, whereas Gaussian-like distributions are considerably shifted in opposite $y$-directions for $\varepsilon=\pm 0.014$. Second, we measured the dependence of the transverse shift of the centroid of the transmitted beam (for $\varepsilon= \pm 0.014$) on the tilt angle $\theta$. This is plotted in Fig.~3(b), together with the theoretical prediction, Eq.~(\ref{E7}). One can clearly see a good agreement between theory and experiment, as well as the considerable role of the nonlinear weak-measurement correction in Eq.~(\ref{E7}) for small $\theta$.


In conclusion, we have examined the SHEL and beam shifts at a tilted linear-dichroic plate (polarizer). Although the SHEL at a tilted polarizer was previously associated with the so-called ``{\it geometric} SHEL'' \cite{Aiello2009,Korger2014}, we have shown that this phenomenon represents an example of the regular SHEL at a tilted anisotropic interface \cite{Bliokh2016}, and it is fully described by the standard equations underlying optical beam shifts at various interfaces \cite{Bliokh2013,Dennis2012,Gotte2012,Toppel2013}.  Furthermore, we have employed the ``weak-measurement'' method to strongly amplify and measure the SHEL at a tilted polarizer by transforming it into the {\it angular} beam shift \cite{Hosten2008, Qin2009, Zhou2012, Gorodetski2012, Dennis2012, Gotte2012, Toppel2013, Jayaswal2014, Goswami2014, Bliokh2016}. Our results once again demonstrate the generic and universal character of the SHEL at various optical interfaces.

This work was partially supported by MURI Center for Dynamic Magneto-Optics via the Air Force Office of Scientific Research (AFOSR) (FA9550-14-1-0040), Army Research Office (ARO) (Grant No. W911NF-18-1-0358), Asian Office of Aerospace Research and Development (AOARD) (Grant No. FA2386-18-1-4045), Japan Science and Technology Agency (JST) (Q-LEAP program, and CREST Grant No. JPMJCR1676), Japan Society for the Promotion of Science (JSPS) (JSPS-RFBR Grant No. 17-52-50023, and JSPS-FWO Grant No. VS.059.18N), RIKEN-AIST Challenge Research Fund, the John Templeton Foundation, SERB India-TARE project (File No. TAR/2018/000552), Science Engineering Research Board (SERB), India, and the Australian Research Council.
\bigskip



\bibliographyfullrefs{sample}

\ifthenelse{\equal{\journalref}{aop}}{%
\section*{Author Biographies}
\begingroup
\setlength\intextsep{0pt}
\begin{minipage}[t][6.3cm][t]{1.0\textwidth} 
  \begin{wrapfigure}{L}{0.25\textwidth}
    \includegraphics[width=0.25\textwidth]{john_smith.eps}
  \end{wrapfigure}
  \noindent
  {\bfseries John Smith} received his BSc (Mathematics) in 2000 from The University of Maryland. His research interests include lasers and optics.
\end{minipage}
\begin{minipage}{1.0\textwidth}
  \begin{wrapfigure}{L}{0.25\textwidth}
    \includegraphics[width=0.25\textwidth]{alice_smith.eps}
  \end{wrapfigure}
  \noindent
  {\bfseries Alice Smith} also received her BSc (Mathematics) in 2000 from The University of Maryland. Her research interests also include lasers and optics.
\end{minipage}
\endgroup
}{}

\end{document}